\documentclass[10pt,aps,pra,tightenlines,twocolumn]{revtex4}

\usepackage{amssymb}
\usepackage{epsf}
\usepackage[hypertex]{hyperref}

\def\diag{\mathop{\rm diag}}

\def\tr{\mathop{\rm Tr}}

\def\1{\openone}
\def\x{\sigma^{x}}

\def\L{\Lambda}
\def\Rv{\textbf{R}}
\def\G{\Gamma}

\def\<{\left\langle}
\def\>{\right\rangle}
\def\X{{\rm X}}
\def\Y{{\rm Y}}

\begin{document}

\title{Soft-Pulse Dynamical Decoupling with Markovian Decoherence}
\author{Leonid P. Pryadko}
\affiliation{Department of Physics \& Astronomy, University of California,
  Riverside, California, USA}
\author{Gregory Quiroz}
\affiliation{Department of Physics, University of Southern California, Los
  Angeles, California, USA}
\begin{abstract}
  We consider the effect of broadband decoherence on the
  performance of refocusing sequences, having in mind applications of
  dynamical decoupling in concatenation with quantum error correcting
  codes as the first stage of coherence protection.  Specifically, we
  construct cumulant expansions of effective decoherence operators for
  a qubit driven by a pulse of a generic symmetric shape,
  and for several sequences of $\pi$- and $\pi/2$-pulses.  While, in
  general, the performance of soft pulses in decoupling sequences in
  the presence of Markovian decoherence is worse than that of the
  ideal $\delta$-pulses, it can be substantially improved by shaping.
\end{abstract}
\maketitle

\section{Introduction}
Dynamical
decoupling\cite{slichter-book,hodgkinson-emsley-2000,vandersypen-2004}
(DD) can be very effective in protecting 
coherence of a quantum system against low-frequency
environment\cite{%
  haeberlen-waugh-1968,%
  haeberlen-waugh-1969,%
  pines-waugh-1972,%
  pines-waugh-1974,%
  viola-lloyd-1998,% 
  vitali-tombesi-1999,%
  viola-knill-lloyd-1999,viola-knill-lloyd-1999B,%
  viola-knill-lloyd-2000,%
  vitali-tombesi-2002,%
  uchiyama-aihara-2002,%
  byrd-lidar-2003,%
  shiokawa-lidar-2004,%
  faoro-viola-2004,facchi-nakazato-2004,%
  sengupta-pryadko-ref-2005,%
  khodjasteh-Lidar-2005,%
  pryadko-sengupta-kinetics-2006,chen-2007,%
  Uhrig-2007,Hodgson-2008,Pasini-2008}. This, combined with low
resource requirement, makes it attractive as the first-level coherence
protection technique, in combination with quantum error correcting
codes\cite{Byrd-2003,Viola-2002} (QECC).  For such a combined
decoherence protection technique to work universally, the performance
of the DD should not be reduced in the presence of fast environment
modes whose effect on coherence is to be dealt with by QECC.

Previously, the effects of broadband noise on DD were analyzed in a
number of publications, with the primary target being the $1/f$ or
telegraph noise\cite{shiokawa-lidar-2004,faoro-viola-2004,%
  mottonen-2006,kuopanportti-2008,%
  Cywinski-2008,Rebentrost-2009}.  However, the
effectiveness of refocusing sequences was mostly studied numerically,
apart from special exactly-solvable cases\cite{Uhrig-2008,Cywinski-2008}.
No attempt has been made to investigate general properties and
limitations of decoupling under these conditions.  Certainly, there
were no analytical studies of effects of pulse shaping on decoupling
in the presence of broadband decoherence.

\begin{figure}[htbp]
  \centering
  \epsfxsize=2.6in
  \epsfbox{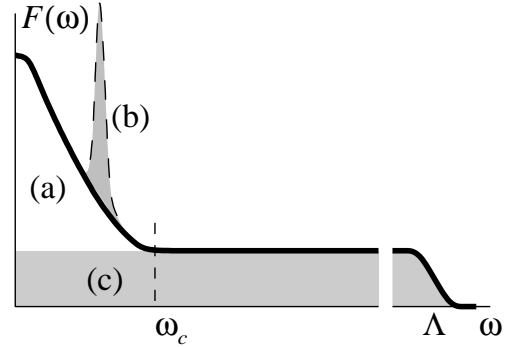}
  \caption{Assumed spectral function of a generic oscillator bath (schematic).
    Dynamical decoupling can be very effective in preserving coherence
    in the presence of (\textbf{a}) featureless low-frequency
    bath\cite{pryadko-sengupta-kinetics-2006} and also (\textbf{b})
    sharp resonances\cite{pryadko-quiroz-2007}.  In this work we
    consider the effect of Markovian noise (\textbf{c}) on DD of a
    single qubit.  We assume $F(\omega)=$const up to a large cut-off
    frequency $\omega_{\rm max}\equiv \Lambda$, so that the qubit
    dynamics associated with the modes (\textbf{c}) can be modeled by
    the Linblad equation~(\protect\ref{eq:Lindblad}).  When the
    high-frequency spectrum is not flat but is, e.g., a power law
    $\propto 1/\omega^{\alpha}$ as in
    Refs.~\protect\onlinecite{Uhrig-2008,Cywinski-2008}, the
    decomposition into low-frequency ({\bf a}) and Markovian ({\bf c})
    parts becomes approximate, with the Markovian decoherence rates
    $\tilde\gamma$ determined by $F(\omega)$ near $\omega=\Omega\equiv
    2\pi/\tau$ determined by the sequence duration $\tau$.}
  \label{fig:spectr}
\end{figure}

In this work, we concentrate on the dynamical decoupling (DD) of a
single qubit (spin) in the presence of a broad-spectrum oscillator
bath [Fig.~\ref{fig:spectr}].  We first consider the effect of \textbf{the pulse shape}
in the special case where the external spin couplings can be
described by a combination of a Markovian noise (described by the
Lindblad decoherence operators) and a time-independent magnetic field.
This basic problem is similar in importance to the canonical
pulse-shaping for a nuclear spin in the presence of a chemical shift.
For a qubit driven by an arbitrarily-shaped pulse, we construct an
analytical expansion of the average decoherence operator, an analogue
of the average Hamiltonian
expansion\cite{Waugh-Huber-Haeberlen-1968,waugh-wang-huber-vold-1968}
but for the qubit density matrix evolution.  Analyzing the first two
terms of such expansions for several decoupling sequences, we
formulate the conditions necessary for improved coherence of the
qubit, and construct numerically the pulse shapes that satisfy these
conditions.  We then compare the performance of the obtained pulse
shapes with ``hard'' pulses and the conventional first- and
second-order nmr-style self-refocusing pulses in several decoupling
sequences both analytically and numerically.  For numerics, we model
the oscillator bath as a combination of classical correlated gaussian
noise and the Markovian noise [Fig.~\ref{fig:spectr}(a) and
  \ref{fig:spectr}(c) respectively].

We analyze two kinds of decoupling sequences, the usual decoupling
sequences utilizing $\pi$-pulses, and the sequences using composite
pulses constructed of three $\pi/2$ pulses similar to those used in
the NMR WaHuHa experiment\cite{Waugh-Huber-Haeberlen-1968}.  While the
former sequences interact relatively little with the decoherence
operators ($\delta$-pulses leave the diagonal parts of the decoherence
operators invariant), the latter ones are constructed to symmetrise
the decoherence operator between all three channels.  Such a
redistribution 
seeks to minimize the detrimental effects of the non-Hermitian evolution
caused by the decoherence operators on higher-level control sequences.
 
\section{Model}
\subsection{Hamiltonian}
A system of individually-controlled qubits in the presence of an oscillator
thermal bath can be described by the following idealized Hamiltonian
\begin{equation}
  \label{eq:ham}
  H=H_C(t)+H_S+H_{SB}+H_B,
\end{equation}
where the control Hamiltonian 
\begin{equation}
  H_C(t)={1\over2}\sum_{n,\alpha}
  V_n^\alpha(t)\,\sigma_n^\alpha\label{eq:ham-control},  
\end{equation}
the system Hamiltonian with one- and few-qubit terms,
\begin{equation}
  \label{eq:ham-system}
  H_S={1\over2}\sum_{n,\alpha} B_n^\alpha\sigma_n^\alpha+
  {1\over4}\sum_{n,n',\alpha,\beta}J_{n,n'}^{\alpha\beta}\,
  \sigma_n^\alpha\sigma_{n'}^\beta+\cdots,
\end{equation}
the linear oscillator bath, 
\begin{equation}
  \label{eq:ham-bath}
  H_B=\sum_\mu\omega_\mu a_\mu ^\dagger a_\mu\,,
\end{equation}
 and the bath-coupling Hamiltonians, 
\begin{equation}
  \label{eq:ham-system-bath}
  H_{SB}={1\over2}\sum_{n,\alpha} \hat b_n^\alpha \sigma_n^\alpha
  +
  {1\over4}\sum_{n,n',\alpha,\beta}\hat {\j}_{n,n'}^{\alpha\beta}\,
  \sigma_n^\alpha\sigma_{n'}^\beta+\cdots . 
\end{equation}
Here $\sigma_n^\alpha$, $\alpha=x,y,z$, are the Pauli matrices for
$n$-th qubit, $a_\mu$, $a_\mu^\dagger$ are the annihilation and
creation operators for the oscillators of the bath, and the operators
$\hat b_n^\alpha$, ${\hat{\j}}_{nn'}^{\alpha\beta}$ describe the various
couplings with the bath.  A linear in phonons coupling, e.g., for the
single-spin term can be written as
\begin{equation} 
 \hat b_n^\alpha=\sum_\mu{f_{n\mu}^\alpha a_\mu+\bar f_{n\mu}^\alpha
    a_\mu^\dagger\over (2m_\mu\omega_\mu)^{1/2}}.
  \label{eq:single-spin-bath-coupling}  
\end{equation}
In the limit where the oscillator modes have a continuous spectrum, the
kinetics of the decoherence effects is determined by the set of spectral
coupling functions (matrices), e.g., for the single-qubit coupling, 
\begin{equation}
  \label{eq:spectral-matrix}
  F_{n\alpha, n'\beta}(\omega)\equiv {\pi\over 2}\sum_\mu {f_{n\mu}^\alpha\bar
    f_{n'\mu}^\beta\over m_\mu \omega_\mu}\delta(\omega-\omega_\mu).
\end{equation}
A schematic plot of a single representative component of this
function is shown in Fig.~\ref{fig:spectr}.  

For simplicity, we assume a combination of weak bath coupling and
finite temperature $T$, so that the (generally non-Markovian) master
equation\cite{konstantinov-perel-1960,davies-1974,dykman-1978,%
  alicki-1989,pryadko-sengupta-kinetics-2006} in the leading order
(``Born approximation'') is satisfied for the combination of
$H_S+H_C(t)$.  For the case of a single qubit, which is primarily
discussed in this work,
\begin{eqnarray}
  \label{eq:master-eq-general}
  \dot\rho&=&\! -i[H_S+H_C(t),\rho]-{1\over4}\!\int_0^t \! dt'\,
  \mathcal{F}^{\alpha\beta}_1(t-t')[\sigma^\alpha,[\sigma^\beta,\rho]]
  \nonumber 
  \\
  & & \!  -{i\over4} \! \int_0^t\! dt'\,
  \mathcal{F}^{\alpha\beta}_2(t-t')[\sigma^\alpha,\{\sigma^\beta,\rho\} ],
\end{eqnarray}
where the dissipation kernel
$\hat\mathcal{F}_1(t)+i\hat\mathcal{F}(t)\equiv \hat \mathcal{F}(t)$
is defined as 
\begin{equation}
  \label{eq:kernel}
  \hat \mathcal{F}(t)  =\int_0^\infty {d\omega\over \pi} [\hat F(\omega)
   (n_\omega+1)e^{i\omega t}+\hat F^{*}(\omega) n_\omega e^{-i\omega t}],
\end{equation}
 $n_\omega\equiv \exp(\beta\omega)-1$ is the oscillator
occupation number, and $\beta\equiv \hbar/T$.

Dynamical decoupling in the presence of a low-frequency bath
[Fig.~\ref{fig:spectr}(\textbf{a}),(\textbf{b})] was analayzed in
Refs.~\cite{pryadko-sengupta-kinetics-2006,pryadko-quiroz-2007}.  General
conclusion is that a \emph{carefully-designed} sequence with the period
$\tau\equiv 2\pi/\Omega$ provides an excellent decoherence protection as long
as the adiabaticity condition $\Omega\agt \omega_c$ is satisfied.  Here
$\omega_c$ is the bath cut-off frequency such that all
$\omega_\mu<\omega_c$ in Eq.~(\ref{eq:ham-bath}).

More precisely, the analysis in
Ref.~\onlinecite{pryadko-sengupta-kinetics-2006} was done for a
featureless low-frequency bath [Fig~\ref{fig:spectr}(\textbf{a})] in
the approximation of non-Markovian quantum kinetic (master) equation.
The approach was to design a decoupling sequence effective for a
closed system with the ``frozen'' bath, where each term in the
bath-coupling Hamiltonian $H_{SB}$ [Eq.~(\ref{eq:ham-system-bath})] is
replaced by the corresponding term with a non-zero $c$-number
coefficient in Eq.~(\ref{eq:ham-system}) [e.g., $\hat b_n^\alpha\to
  B_n^\alpha$, $\hat {\j}_{n,n'}^{\alpha,\beta}\to J_{n,n'}^{\alpha,\beta}$,
  etc., with the matching indices.] The corresponding unitary
evolution operator $U(t=\tau)$ after one decoupling period can be
characterized by the decoupling order $K$, the number of terms in the
Magnus (cumulant) expansion in powers of $H_S$ which are suppressed
identically.  Such a suppression can be expressed in terms of certain
algebraic conditions which were used in the analysis of the solutions
of the full master equation, order-by-order in powers of the
adiabaticity parameter $\omega_c/\Omega$.  At $K=1$, the single-phonon
$T_1$ processes are completely suppressed as long as $\omega_c<\Omega$
(this can also be seen by analyzing the absorption spectra of the
driven system\cite{kofman-kurizki-2004}).  Additionally, since the
bath coupling is modulated at the frequency $\Omega$, the effective
dephasing rate is reduced by the factor of
$\omega_c/\Omega$\cite{pryadko-sengupta-kinetics-2006}.  At $K=2$, in
many cases (e.g., for a single qubit), dephasing rate is suppressed
faster than any power of $\omega_c/\Omega< 1$: terms of every order in
the corresponding expansion are
suppressed\cite{pryadko-sengupta-kinetics-2006}.  

The description in terms of the master equation fails in the presence
of sharp resonances [Fig.~\ref{fig:spectr}(\textbf{b})]; the
corresponding modes have to be included in the Hamiltonian of the
system and the controlled dynamics
re-analyzed\cite{pryadko-quiroz-2007}.  Compared to qubits-only
systems, presence of oscillators in $H_S$ can generate some additional
terms in the Magnus expansion of the unitary evolution operator
$U(\tau)$ for the closed system which requires more careful sequence
design.  When this is done, only the effective, DD-renormalized
coupling to such modes matters.  Non-equilibrium effects like mode
heating do not occur as long as this renormalized coupling is small
compared to either the width of the spectral peak or the corresponding
frequency bias\cite{pryadko-quiroz-2007}.

\subsection{Markovian decoherence}
We now consider the dynamical decoupling in the presence of broadband
bath modes, see Fig.~\ref{fig:spectr}(\textbf{c}), assuming that the
corresponding inverse correlation time $\Lambda\equiv \omega_{\rm
  max}$ is much larger than the control bandwidth which determines the
pulse rate which can be implemented for the system.  Out of the master
equation (\ref{eq:master-eq-general}), we separate a
part $\mathcal{D}(\rho)$, Markovian on the time scale slow compared to
$\omega_{\rm max}$, and assume that the remaining relatively slow
modes are limited to frequencies $\omega\le \omega_c\ll\omega_{\rm
  max}$.  Most generally, Markovian evolution implies the dissipator
in the Lindblad form\cite{lindblad-76}, 
\begin{equation}
  \label{eq:dissipator}
  \mathcal{D}(\rho)\equiv  {1\over2}\sum_{j} 
  \left(\bigl[\L_j\rho,\L^{\dagger}_j\bigr]+
  \bigl[\L_j,\rho\L^{\dagger}_j\bigr]\right)
\end{equation}
where $\rho$ is the density matrix and $\Lambda_j$ are the decoherence
operators.  Such a form can also be recovered from
Eq.~(\ref{eq:kernel}) in the Markovian limit where the time-dependence
of the kernel $\hat \mathcal{F}(t)$ becomes $\delta$-function-like.
We note that this separation of the approximately Markovian part of
the bath, Eq.~(\ref{eq:dissipator}), does not imply the consitions of
high temperature or weak coupling combined with adiabaticity which are
usually necessary for the applicability of the Markovian master
equation\cite{davies-1974,alicki-1989,alicki-2006}.  

For example, with flat part of $F(\omega)\approx{\rm const}$ as in
Fig.~\ref{fig:spectr}(c), we could set the low-frequency cut-off
$\omega_c\agt \beta^{-1}$ and include the effect of
$(2n_\omega+1)\mathcal{F}_1(\omega)$ in Eq.~(\ref{eq:kernel}) in the
non-Markovian part of the bath kernel.  This is precisely the case
solved numerically in Sec.~\ref{sec:simulations} where the slow bath
modes are replaced by correlated classical noise as appropriate for
$\hbar\omega\alt T$.

To start, we will follow the approach of
Ref.~\onlinecite{pryadko-sengupta-kinetics-2006}, and begin by
analyzing the dynamics of the system with ``frozen'' dynamics of the
slow modes in Fig.~\ref{fig:spectr}(\textbf{a}).  That is, we will
assume that the decoupling frequency $\Omega$ is large compared to
$\omega_c$, so that any bath couplings can be approximated by
appropriate terms in $H_S$.  Then, the effect of the fast
environmental modes [Fig.~\ref{fig:spectr}(\textbf{c})] on dynamical
decoupling can be analyzed by considering the driven Lindblad
equation,
\begin{equation}
\dot{\rho}=-i\left[H_C(t)+H_S,\rho\right]+
\mathcal{D}(\rho).
\label{eq:Lindblad}
\end{equation}

For a single qubit, parametrize the dissipator in terms of a Hermitian
non-negative-definite rate matrix $\tilde\gamma_{\alpha\beta}$,
\begin{equation}
  \mathcal{D} (\rho)={1\over4}\sum_{\alpha,\beta=x,y,z}
  {\tilde\gamma_{\alpha\beta}}
  \left(\bigl[\sigma^\alpha\rho,\sigma^\beta\bigr]+
    \bigl[\sigma^\alpha,\rho\sigma^\beta\bigr]\right).
  \label{eq:decoherence}
\end{equation}
It is convenient to separate the symmetric real and antisymmetric
imaginary parts of the rate matrix,
\begin{equation}
  \tilde\gamma_{\alpha\beta}=\gamma_{\alpha\beta}
  +i\epsilon_{\alpha\beta\gamma}\mathcal{R}_\gamma.
\label{eq:gamma-decomposition}
\end{equation}
Then, if we
rewrite the density matrix in terms of the Bloch vector $\Rv\equiv
\Rv(t)$, 
\begin{equation}
  \rho(t)=\frac{1}{2}\left(\openone+\Rv(t)\cdot\vec{\sigma}\right),
  \quad R^2\le 1, 
  \label{eq:bloch}
\end{equation}
the equation for spin kinetics in the presence of the single-spin
control~(\ref{eq:ham-control}) and 
system~(\ref{eq:ham-system}) 
Hamiltonians becomes
\begin{eqnarray}
  \label{eq:Revol}
  \dot \Rv&=&[{\bf V}(t)\times\Rv]-\hat\Gamma\, \Rv-\vec\mathcal{R},\\
   -\hat\Gamma\, \Rv & \equiv& [{\bf
  B}\times\Rv]+(\hat\gamma-\openone\tr{\hat\gamma})\,\Rv(t).
\label{eq:Gamma}
\end{eqnarray}
The terms with $\mathbf{V}(t)$ and $\mathbf{B}$ describe precession in the
effective magnetic fields of the control and the system Hamiltonians, the term
with $\hat\gamma$ describes the coherence loss due to Markovian bath, while
the last term in Eq.~(\ref{eq:Revol}) ensures the correct equilibrium value of
$\Rv$ in the absence of control.

If we work in the basis associated with the dissipator, the rate matrix
$\hat\gamma$ is diagonal.  In particular, for the case of nuclear magnetic
resonance (or any physical qubit with large energy difference between the
levels), we have $\gamma_{xx}=\gamma_{yy}=\gamma$, $\gamma_{zz}=\gamma_\phi$,
and the matrix~(\ref{eq:Gamma}) takes the form 
\begin{equation}
  \label{eq:Gamma-phase}
  \hat\Gamma=\left(
  \begin{array}[c]{ccc}
    \gamma+\gamma_{\phi} & B_z & -B_y\\
    -B_z &     \gamma+\gamma_{\phi} & B_x\\
    B_y & -B_x & 2\gamma
  \end{array}\right).
\end{equation}
These decay rates correspond to the usual coherence times $T_1^{-1}=2\gamma$,
$T_2^{-1}=\gamma+\gamma_\phi$.  The corresponding vector $\mathcal{R}$ has
only one component, $\mathcal{R}_z\ge0$.
In the special case $\gamma_\phi=\gamma$, the
rate matrix is proportional to the identity matrix,
$\hat\gamma=\gamma\,\openone$.

We emphasize again that, while the analytical analysis is done
assuming time-independent ``magnetic field'' ${\bf B}$ in
Eq.~(\ref{eq:Gamma-phase}), the decoupling accuracy for obtained pulse
shapes is also verified numerically in Sec.~\ref{sec:simulations} with
the components $B_\mu(t)$ taken as a classical time-dependent Gaussian
correlated noise.

\subsection{Interaction representation \& Magnus expansion}

In dynamical decoupling, it is the control Hamiltonian $H_C(t)$ that
dominates the dynamics.  The evolution associated with the system Hamiltonian
$H_S$ [Eq.~(\ref{eq:ham-system})] or the dissipator $\mathcal{D}(\rho)$
[Eq.~(\ref{eq:dissipator})] is much slower.  Thus, it is convenient to
consider the dynamics in the interaction representation with respect to
$H_C(t)$.  To this end, we introduce the non-perturbed unitary evolution
operator $U_0 \equiv U_0(t)$,
\begin{equation}
  U_0(t)=T_t \exp \Biglb(-i\int_0^t dt'\,H_C(t')\Bigrb),
\label{eq:unitary-bare}
\end{equation}
where $T_t$ represents the usual time-ordering operator, and the corresponding
orthogonal spin-rotation matrix, $\hat Q_0\equiv \hat Q_0(t)$, can be computed
as follows
\begin{equation}
  \label{eq:orhogonal-bare}
  U_0\sigma^\alpha U_0^\dagger =Q_0^{\alpha\beta} \sigma^\beta.
\end{equation}
The components $ Q_0^{\alpha\beta}$ satisfy the
Bloch equation
\begin{equation}
 \dot
  Q_0^{\alpha\beta}(t)=e^{\alpha\gamma\delta}\,
  {V}^\gamma(t)\,Q_0^{\delta\beta}(t),\quad \hat Q_0(0)=\openone.
  \label{eq:bare-evolution}
\end{equation}

If we write the solution of the uniform version of Eq.~(\ref{eq:Revol}) (i.e.,
with $\vec\mathcal{R}=0$) in terms of the full evolution matrix $\hat Q\equiv
\hat Q(t)$,
\begin{equation}
  \label{eq:uniform-sol}
  \mathbf{R}(t)=\hat Q(t)\, \mathbf{R}_0,
\end{equation}
we can introduce the decomposition $\hat Q=\hat Q_0 \, S$, where the
matrix $S$ defines the slow evolution of the Bloch vector in the rotating
frame defined by the control fields.  The equation for $S$ reads
\begin{equation}
  \dot S=-\hat \Gamma(t)\,S,\quad \hat\Gamma(t)\equiv \hat Q_0^\mathrm{t}(t)\,
  \hat \Gamma\,\hat Q_0(t), 
\label{eq:Sevol}
\end{equation}
where $\hat\Gamma(t)$ is the evolution operator~(\ref{eq:Gamma}) in the
interaction representation.  The formal solution of Eq.~(\ref{eq:Sevol})
can be again written as a time-ordered exponent,
\begin{equation}
  S(t)=T_t\exp\Biglb(-\int^{t}_{0}dt' \,\hat\Gamma(t')\Bigrb).
  \label{eq:Sformal}
\end{equation}

For a periodic control field, $H_c(t+\tau)=H_c(t)$, such that the
zeroth-order rotation matrix is also periodic, $Q_0(t+\tau)=Q_0(t)$, the
time-ordered exponent can be evaluated in terms of the average decoherence
operator $\overline\G$, an analogue of the average
Hamiltonian\cite{Waugh-Huber-Haeberlen-1968,waugh-wang-huber-vold-1968}:
\begin{equation}
S(n\tau)=e^{-n\tau \,\overline\G},\quad 
\overline\G\equiv \overline\G^{(0)}+\overline\G^{(1)}+\cdots,
\label{eq:Ggeneral}
\end{equation}
where
\begin{eqnarray}
  \overline\G^{(0)}&=&{1\over \tau}\int^{\tau}_{0}dt_1\tilde\Gamma(t_1),
  \label{eq:G0}\\
  \overline\G^{(1)}&=&-\frac{1}{2\tau}\int^{\tau}_{0}dt_2
  \int^{t_2}_{0}dt_1[\hat\Gamma(t_2),\hat\Gamma(t_1)],
  \label{eq:G1}
\end{eqnarray}
etc.  Generally, the term $\overline\G^{(k)}$, $k>0$, of this
expansion contains a $(k+1)$-fold integration of commutators of the
rotating-frame decoherence operator $\hat\G(t_j)$; it scales as
$\propto \tau^{k}$.  Trace of a commutator is zero, thus
$\tr{\overline\G}^{(k)}=0$ for all $k>0$; this implies 
\begin{equation}
  \tr {\overline\G}=\tr \overline\G^{(0)}=\tr \hat \Gamma=2\tr\hat\gamma.
  \label{eq:trace}
\end{equation}
Generally, the trace-conserving modifications of the effective decoherence
operator correspond to (some degree of) symmetrization between the decoherence
channels.  More precisely, the real part of the smallest eigenvalue of
$\overline\Gamma$ cannot be smaller than that of $\hat\Gamma$.  This can be
seen by separating out the trivial part of $\hat\Gamma$ proportional to
identity matrix, which corresponds to uniform decoherence for all channels and
remains unchanged in the interaction representation~(\ref{eq:Sevol}).

\subsection{Coherence loss in controlled system}
\label{sec:coherence-loss}
As would be expected on general grounds, Eq.~(\ref{eq:trace}) implies that the
DD cannot eliminate the Markovian dissipation; at best we can hope to
eliminate the off-diagonal terms of matrix $\hat\Gamma$ and redistribute the
rates over the decoherence channels.  For example, for an NMR qubit
decoherence model, Eq.~(\ref{eq:Gamma-phase}), redistribution (symmetrization)
over all three channels would lead to
\begin{equation}
  \overline\G={2\over 3}(2\gamma+\gamma_\phi)\,\openone.
\label{eq:Gamma-symmetric}  
\end{equation}
With $\gamma\ll\gamma_\phi$, this is equivalent to a 33\% reduction of the
maximum decoherence rate, or a 50\% increase of the coherence time measured by
the fidelity minimized over the initial conditions.  

We should note that the quoted estimate for the maximum decoherence
rate improvement does not take into account the antisymmetric part of
the rate matrix $\tilde\gamma$ [Eq.~(\ref{eq:gamma-decomposition})],
or, equivalently, the non-uniform term $\vec\mathcal{R}$ in
Eq.~(\ref{eq:Revol}).  In the absence of control fields, this term is
responsible for asymptotic thermal-equilibrium orientation of the
spin; it is small and can be ignored at sufficiently high temperatures
(e.g., room-temperature NMR), but it can be large at temperatures
small compared to the qubit level difference $\Delta
E_{01}=|\mathbf{B}|$.

We can drop the non-uniform term $\vec\mathcal{R}$ in the analysis of
decoherence since it does not affect the evolution of the \emph{average}
fidelity, that is, the fidelity averaged by the initial conditions,
$\bar\mathcal{F}(t)\equiv  \tr\langle\rho(0)\rho(t)\rangle_0$.  Indeed,
Eq.~(\ref{eq:Revol}) is a set of linear differential equations for the
components of the Bloch vector ${\bf R}(t)$; the solution with given initial
conditions $\mathbf{R}(0)\equiv \mathbf{R}_0$ can be written as a sum of the
solution~(\ref{eq:uniform-sol}) of the uniform equation with the same initial
condition, and that of the non-uniform equation but with zero initial
condition.  In the expression for the average fidelity,
\begin{equation}
  \bar\mathcal{F}(t) ={1\over
    2}(\openone+\langle \mathbf{R}_0\cdot \mathbf{R}(t)\rangle_{\mathbf{R}_0})
  \label{eq:av-fid}
\end{equation}
the solution of the uniform equation is bilinear in the components of
$\mathbf{R}_0$ while the non-uniform part is linear; the averaging over all
directions of $\mathbf{R}$ leaves only the bilinear part,
\begin{equation}
  \label{eq:av-fid-evaluated}
  \bar \mathcal{F}(n\tau)  ={1\over2}+{1\over6}\tr \hat Q(n\tau)
  ={1\over 2}+{1\over 6}\tr 
  e^{-n\tau\overline\Gamma},
\end{equation}
where we assumed the zeroth-order rotation matrix to be periodic, $\hat
Q_0(n\tau)=\openone$, and used the representation (\ref{eq:Ggeneral}) in
terms of the average decoherence operator $\overline\Gamma$.

The obtained expression~(\ref{eq:av-fid-evaluated}), along with the
properties of the average decoherence operator $\overline\G$ discussed
in the previous section [see Eq.~(\ref{eq:trace})], imply that the
average fidelity cannot be improved by means of dynamical decoupling
beyond eliminating the effect of the system Hamiltonian $H_S$
[Eq.~(\ref{eq:ham-system})].  With $\mathbf{B}=0$, this follows
directly from the convexity of exponentials in
Eq.~(\ref{eq:av-fid-evaluated}) if one works in the basis where the
average decoherence matrix
$\bar\gamma=\diag(\gamma_1,\gamma_2,\gamma_3)$,
$0\le\gamma_1\le\gamma_2 \le\gamma_3$.  In particular, the maximum
possible fidelity with given $\tr\hat\gamma$ corresponds to only one
non-zero component, $\gamma_3=\tr\hat\gamma$.  In this case two (or
any even number) qubit errors compensate each other.  On the other
hand, the symmetric case $\gamma_1=\gamma_2=\gamma_3= \tr\hat\gamma/3$
in the absence of error correction corresponds to the fidelity
minimum.

We should also note that a symmetrization of decoherence rates over channels
[complete as in Eq.~(\ref{eq:Gamma-symmetric}), or partial] would prohibit the
use of special QECCs designed for strongly asymmetric error rates between the
channels\cite{Gourlay-2000,Ioffe-Mezard-2007,Evans-2007,sarvepalli-2009,%
  Aliferis-Preskill-2008,Stephens=2008}.  

In this work we assume that the benefits of dynamical decoupling (due to
reduced decoherence coming from the slow degrees of freedom) outweigh the
detrimental effects related to symmetrization of the coherence rates.

\section{Pulse and sequence characterization}
\subsection{Hard pulses}
Let us now use the obtained formalism to analyze controlled dynamics.  We
start with hard pulses, with the $\delta$-function centered at the middle of
the interval of duration $\tau_p$.  For example, for a $(\phi_0)_x$ pulse
(rotation angle $\phi_0$ with respect to $x$-axis), the non-zero control field
in Eq.~(\ref{eq:ham-control})
can be written as 
\begin{equation}
V^x(t)=\phi_0\delta(t-\tau_p/2),\; 0<t<\tau_p.\label{eq:delta-pulse-x}
\end{equation}
For the special case $\phi_0=\pi$ (pulse $\pi_x$), the corresponding rotation
matrix~(\ref{eq:bare-evolution}) is 
\begin{equation}
  \label{eq:bare-rotation-matrix-pi}
  \hat Q_0
  (t)=\left\{
    \begin{array}[c]{cc}
      \openone,&0<t<\tau_p/2;\\
      \diag(1,-1,-1),&\tau_p/2<t<\tau_p.
    \end{array}\right.
\end{equation}
Over the second half of the interval, the components $\Gamma_{xy}$ and
$\Gamma_{xz}$ are inverted; as a result in the leading order 
\begin{equation}
  \label{eq:decoherence-matrix-pi}
  \overline\G^{(0)}(\pi_x)=\left(
    \begin{array}[c]{ccc}
      \gamma_{yy}+\gamma_{zz} & & \\
      & \gamma_{zz}+\gamma_{xx} & B_x-\gamma_{yz}\\
      & -B_x-\gamma_{yz} & \gamma_{xx}+\gamma_{yy}
    \end{array}\right).
\end{equation}
The expression for $\overline\G^{(1)}(\pi_x)$ is too complicated to quote here.
However, for the symmetric sequence $\textbf{2s}\equiv \X\X$ of
$X\equiv \pi_x$
 pulses (which is the repeated part of the
Carr-Purcell sequence), the first-order average decoherence operator
$\overline\G^{(1)}(\mathbf{2s})=0$, while the zeroth-order term
$\overline\G^{(0)}(\mathbf{2s})$ is the same as for a single pulse given by
Eq.~(\ref{eq:decoherence-matrix-pi}).  

Deriving similar expressions for a $\pi_y$ pulse, we can easily
construct the expressions for standard two-dimensional decoupling
sequences of hard $\pi$-pulses.  In particular, the
group-averaging\cite{Zanardi-1999,viola-knill-lloyd-1999} 4-pulse
sequence $\textbf{4p}\equiv \X\Y\bar\X\Y$ results in cancellation of
all off-diagonal terms to leading order,
\begin{equation}
  \label{eq:hard-4p}
  \overline\G^{(0)}(\mathbf{4p})=\left(
    \begin{array}{ccc}
      \gamma _{{yy}}+\gamma _{{zz}} & 0 & 0 \\
      0 & \gamma _{{xx}}+\gamma _{{zz}} & 0 \\
      0 & 0 & \gamma _{{xx}}+\gamma _{{yy}}
    \end{array}
  \right),
\end{equation}
while the symmetric sequence $\mathbf{8s}\equiv \X\Y\bar\X\Y\Y\bar\X\Y\X$
(symmetrized version of \textbf{4p}) results in cancellation of all
off-diagonal terms
[$\overline\G^{(0)}(\mathbf{8s})=\overline\G^{(0)}(\mathbf{4p})$, see
Eq.~(\ref{eq:hard-4p})] and does not produce any 1st-order corrections,
$\overline\G^{(1)}(\mathbf{8s})=0$.  We note that with the hard pulses, the
contribution to $\overline\G^{(2)}(\mathbf{8s})$ is non-zero already for a
Hamiltonian evolution with $\hat\gamma=0$, as long as both $B_x$ and $B_y$ are
non-zero.

The hard $\pi$ pulses do not modify the structure of the diagonal part
of the evolution matrix $\Gamma$.  A redistribution (symmetrization)
over different decoherence channels can be done if we deliberately
orient the spin along different axes with $\pi/2$-pulses.  In the
absence of the system Hamiltonian ($\mathbf{B}=0$), for NMR
decoherence model~(\ref{eq:Gamma-phase}), the symmetrization can be
achieved with the WaHuHa (MREV-4) NMR
experiment\cite{Waugh-Huber-Haeberlen-1968,Rhim-1973}, a sequence of
$\pi/2$ pulses with the cycle ${\bf 5}=\X \Y \,0\,\bar \Y\bar \X$ of
duration $\tau=5\tau_p$, where each pulse is centered as in
Eq.~(\ref{eq:delta-pulse-x}) and $0$ stands for a free evolution
interval equal to the pulse duration $\tau_p$.
The averaging of the second-rank coupling achieved by this sequence is used to
eliminate dipolar coupling between nuclei.  However, the accuracy
achieved by this sequence is rather sensitive to chemical shifts; the
corrections are present already in the leading order effective evolution
operator.

To achieve both the decoherence symmetrization and the decoupling, we
constructed several pulse sequences based upon the composite pulses
$R_\alpha\equiv\X\bar\Y\X$, $R_\beta\equiv \X\Y\X$, and the inverted
versions $\bar R_\alpha\equiv \bar\X\Y\bar\X$, and $\bar R_\beta\equiv
\bar\X\bar\Y\bar\X$.  Here, $X$ and $Y$ now denote the
  $\pi/2$ pulses applied in the corresponding directions.  The composite
pulses $R_\alpha$ and $R_\beta$ are mutually orthogonal $\pi$-pulses
constructed in such a way that the spin's $x$, $y$, and $z$ axes spend
equal time oriented along the $z$-axis.  As a result, they achieve a
leading-order symmetrization of the NMR decohernce matrix, see
Eq.~(\ref{eq:Gamma-phase}) with $\vec{B}=0$.  The twelve- and
twenty-four pulse sequences, \textbf{12} and \textbf{24} respectively,
are merely the universal decoupling sequences \textbf{4p} and
\textbf{8s} constructed from these composite pulses,
\begin{eqnarray}
  \label{eq:12}
  \mathbf{12}&=&R_\alpha R_\beta \bar R_\alpha R_\beta,\\
  \label{eq:24}
  \mathbf{24}&=&R_\alpha R_\beta \bar R_\alpha R_\beta \,R_\beta \bar R_\alpha
  R_\beta R_\alpha,
\end{eqnarray}
while the \textbf{48}-pulse sequence includes an additional cycle of phase
ramping analogous to that in MLEV-16 and higher order
sequences\cite{Morris-Gibbs-1991},
\begin{equation}
  \label{eq:48}
  \mathbf{48}=R_\alpha R_\beta \bar R_\alpha R_\beta \,
  \bar R_\alpha \bar R_\beta  R_\alpha \bar R_\beta + (\mathrm{reverse}).
\end{equation}
The basic 12-pulse sequence achieves symmetrization but not the decoupling;
the leading-order average decoherence operator reads
\begin{equation}
  \label{eq:decoh-12}
  \overline\G^{(0)}_\mathbf{12}={1\over 6}\left(
    \begin{array}[c]{ccc}
      4(2\gamma+\gamma_\phi) & 2B_y & -B_z\\
      -2B_y &       4(2\gamma+\gamma_\phi) & -B_z\\
      B_z &       B_z &       4(2\gamma+\gamma_\phi). 
    \end{array}
  \right)
\end{equation}
The longer sequence \textbf{24} achieves both the decoupling and
symmetrization of the decoherence operator~(\ref{eq:Gamma-phase}) in
the leading order, i.e., $\overline\G_{\bf 24}^{(0)}$ is given by
Eq.~(\ref{eq:Gamma-symmetric}).  The sequence \textbf{48} achieves
symmetrization to subleading order, i.e., $\overline\G_{\bf 48}^{(0)}$
is given by Eq.~(\ref{eq:Gamma-symmetric}) while $\overline\G_{\bf
  48}^{(1)}=0$.  Moreover, in the experimentally relevant case
$B_x=B_y=0$, the $\overline\G_{\bf 48}^{(2)}$ only has non-zero
elements $\propto B_z^2 (\gamma-\gamma_\phi)\tau_p^2$ along the
diagonal.

\subsection{Finite-duration pulses}
Let us now analyze the performance of discussed sequences when finite-duration
``soft'' pulses are used instead of the ideal hard pulses.  Basically, we
extend the formalism of
Refs.~\onlinecite{pryadko-quiroz-2007,pryadko-sengupta-2008} to the case of
non-Hermitian evolution (\ref{eq:Revol}) with the most general symmetric rate
matrix $\hat\gamma$.

\subsubsection{Average evolution operator for a single pulse}
\label{sec:single-pulse}
We begin with the qubit evolution driven by a one-dimensional pulse
of arbitrary shape,
\begin{equation}
H_c(t)=\frac{1}{2}V(t)\,\x. 
\label{eq:Hc}
\end{equation}
The (zeroth-order) evolution operator due to control field alone is given by
\begin{equation}
U_c(t)=\exp\biglb(-i\x\phi(t)/2\bigrb),\quad \phi(t)=\int^{t}_{0}dt' \,V(t').
\end{equation}
This is just a rotation by angle $\phi(t)$ around the $x$-axis, 
\begin{equation}
        \hat{Q}_0(t)=\left(
  \begin{array}[c]{ccc}
    1 &                 0        & 0\\
    0   & \cos\phi(t)   & \sin\phi(t)\\
    0   & -\sin\phi(t)  & \cos\phi(t)
  \end{array}\right). 
\label{eq:rotation-x}
\end{equation}
With non-zero decoherence matrix $\hat\gamma\neq0$, the evolution operator in
the interaction representation, Eq.~(\ref{eq:Sevol}), will contain terms that
depend both linearly and bi-linearly on the components of the
matrix~(\ref{eq:rotation-x}).  In other words, the time-dependence of
$\hat\Gamma(t)$ will be only through terms proportional to $\cos\phi$,
$\sin\phi$, $\cos2\phi$, and $\sin2\phi$.  The functions of the doubled angle
are specifically due to the Markovian decoherence operators; they were not
present in the analysis of Hamiltonian dynamics in
Refs.~\onlinecite{pryadko-quiroz-2007,pryadko-sengupta-2008}.

For the symmetric pulse shape, 
\begin{equation}
  \label{eq:pulse-symmetry}
  V_x(\tau_p-t)=V_x(t),\quad \phi(\tau_p-t)=\phi_0-\phi(t),
\end{equation}
it is convenient to introduce
the symmetrized angle
\begin{equation}
  \label{eq:symmetrized-angle}
\varphi(t)\equiv\phi(t)-\phi_0/2,\quad 
\varphi(\tau_p-t)=-\varphi(t).
\end{equation}
The averages of time-dependent terms can be then written as follows:
\begin{eqnarray}
  \label{eq:single-avers}
  \langle\cos\phi\rangle=\upsilon \cos{\phi_0\over2},\quad 
  \langle\sin\phi\rangle=\upsilon \sin{\phi_0\over2},\\
  \label{eq:double-avers}
  \langle\cos2\phi\rangle=\upsilon_2 \cos{\phi_0},\quad 
  \langle\sin2\phi\rangle=\upsilon_2 \sin{\phi_0},
\end{eqnarray}
where 
\begin{equation}
  \label{eq:avs-1st}
  \upsilon\equiv \langle\cos\varphi(t)\rangle,\quad
\upsilon_2\equiv \langle\cos 2\varphi(t)\rangle,
\end{equation}
and the pulse-averages are defined simply as 
\begin{equation}
  \label{eq:pulse-average-1}
  \langle f(t)\rangle\equiv {1\over \tau_p}\int_0^{\tau_p} dt\,f(t).
\end{equation}
We see that, to leading order, the pulse-shape dependence of the
average evolution operator $\overline\G^{(0)}$ [Eq.~(\ref{eq:G0})] is
reduced to only two constants.  The parameter $\upsilon$ gives an
effective pulse length to first order; it is equal to zero,
$\upsilon=0$ for ideal $\delta$-inversion pulses (rotation angle
$\phi_0=\pi$), as well as for Hermitian pulses\cite{warren-herm} or
other 1st-order self-refocusing
pulses\cite{sengupta-pryadko-ref-2005,pryadko-sengupta-2008,Pasini-2009}.  The
parameter $\upsilon_2$ vanishes for ideal $\delta$-pulses with the
rotation angle $\phi_0=\pi/2$; for hard inversion pulses
[Eq.~(\ref{eq:delta-pulse-x})] we have $\varphi=\pm\pi/2$, thus $\cos
2\varphi=-1$ throughout the interval, which gives $\upsilon_2=-1$.
For soft pulses with limited amplitude, $|\upsilon_2|<1$. 
The values of the parameters for some other pulse shapes are listed in
Table~\ref{tab:pulses}.
\begin{widetext}
Specifically for a symmetric $\pi_x$ pulse, we get 
\begin{equation}
  \label{eq:one-pi-pulse-evol0}
\overline\G^{(0)}(\pi_x)=
 \left(
\begin{array}{ccc}
 \gamma _{{yy}}+\gamma _{{zz}} & \upsilon(B_y+\gamma_{xz}) & \upsilon(B_z-\gamma_{xy}) \\
 \upsilon(-B_y+\gamma_{xz}) & \gamma
 _{{xx}}+\gamma_{yy}{1+\upsilon_2\over2}+\gamma_{zz}{1-\upsilon_2\over2}
 & 
 v_2  \gamma _{{yz}}+B_x \\
 -\upsilon(B_z+\gamma_{xy}) & v_2 \gamma _{{yz}}-B_x & 
\gamma _{{xx}}+\gamma_{yy}{1-\upsilon_2\over2}+\gamma_{zz}{1+\upsilon_2\over2}
\end{array}
\right),
\end{equation}
which goes over to Eq.~(\ref{eq:decoherence-matrix-pi}) for
$\upsilon=0$, $\upsilon_2=-1$. 
\end{widetext}

\begin{table*}[htbp]
  \centering
  \begin{tabular}[c]{|c|c|c|c|c|c|c|c|c}
    pulse & $\phi_0$ & $\upsilon$ & $\upsilon_2$ & $\alpha/2$ &
    $\alpha_2/2$ & $\zeta$
     & $\zeta_2$  & $\mu$ \\ \hline 
    $\phi_0\delta(t-\tau_p/2)$ & $\phi_0$ & $\cos {\phi_0\over2}$ &
    $\cos \phi_0$ & ${1\over8}\sin\phi_0$ &
    ${1\over8}\sin{2\phi_0}$ & ${1\over4}\sin{\phi_0\over2}$ 
    & ${1\over 4}\sin{\phi_0}{}_{\strut}$&
    ${1\over4}\sin {3\phi_0\over2}$ \\ \hline 
    $\pi\delta(t-\tau_p/2)$& $\pi$ & 0 & $-1$ & $0$ &  $0$ & $1/4$ &
    $0$ & $-1/4$ \\
    $G_{0.01}$\cite{bauer-gauss} &  $\pi$ & $0.0211$ & $-0.9709$ & $0.0104$ &
    $0.000047$ & $0.24996$ & $0.00023$ & $-0.2354$ \\
    $G_{0.10}$\cite{bauer-gauss} & $\pi$ & $0.2107$ & $-0.7086$ & $0.0872$ &
    $0.0047$ & $0.2458$ & $0.0233$ & $-0.1035$ \\
    $S_1$\cite{sengupta-pryadko-ref-2005} &$\pi$ & $0$ & $-0.6135$ & $0.0333$ &
    $0.0393$ & $0.2382$ & $-0.0737$ & $-0.1020$ \\ 
    $S_2$\cite{sengupta-pryadko-ref-2005} &$\pi$ &  $0$ & $-0.6675$ & $0.0250$ & $0.0298$ &
    $0.2414$ & $-0.0557853$ & $-0.1171$ \\ 
    $Q_1$\cite{sengupta-pryadko-ref-2005} &$\pi$ & $0$ & $-0.6761$ & $0$ & $-0.0079$ &
    $0.2399$ & $0.0027$ & $-0.1234$ \\ 
    $Q_2$\cite{sengupta-pryadko-ref-2005} &$\pi$ & $0$ & $-0.7138$ & $0$ & $-0.0065$ &
    $0.2422$ & $0.0022$ & $-0.1342$ \\ 
    $W_{11}(\pi)$ &$\pi$ &  $0$ & $0$ & $0.0511$ & $-0.0039$ &
    $0.1884$ & $-0.1014$ & $0.0353$ \\ 
    $W_{12}(\pi)$ &$\pi$ & $0$ & $0$ & $0.0400$ & $-0.0164$ &
    $0.1904$ & $-0.0871$ & $0.0413$ \\  
    $W_{21}(\pi)$ &$\pi$ & $0$ & $0$ & $0$ & $0.0088$ & $0.0072$ & $0.0677$ & $-0.0093$ \\ 
    $W_{22}(\pi)$ &$\pi$ & $0$ & $0$ & $0$ & $0.0107$ & $0.0634$ & $0.0415$ & $-0.0035$ \\
    $W_{31}(\pi)$ &$\pi$ & $0$ & $0$ & $0$ & $0.00061$ & $0.0436$ & $0$ & $0.0014$ \\
    $W_{32}(\pi)$ &$\pi$ & $0$ & $0$ & $0$ & $0.00046$ & $0.0847$ & $0$ & $0.0146$ \\
    $F_1$ &$\pi$ & $0.0018$ & $0.3307$ & $0.0237$ &
    $-0.01018$ & $0.1134$ & $-0.0260$ & $0.0680$ \\ 
    PKRU$_{1}(\pi)$\cite{Pasini-2009} & $\pi$ & $0$ & $-0.6880$ & $0.0278$ & $0.0368$ &
    $0.2420$ & $-0.0614$ & $-0.1250$ \\ 
    PKRU$_{2}(\pi)$\cite{Pasini-2009} &$\pi$ &$0$ & $-0.1501$ & $0$ & $-0.0078$ & $0$
    & $0.0866$ & $-0.0047$ \\ 
    \hline 
    ${\pi\over2}\delta(t-\tau_p/2)$ & ${\pi/2}$ & $\sqrt2/2$ & $0$
    & $1/8$ & $0$ & $\sqrt2/8$ & $1/4$ & $\sqrt2/8$\\
    $G_{0.01}(\pi/2)$ & $\pi/2$ & $0.7136$ & $0.0211$ &
    $0.1272$ & $0.0104$ & $0.1767$ & $0.2500$ &
    $0.1872$ \\
    $G_{0.10}(\pi/2)$ & $\pi/2$ & $0.7722$ & $0.2107$ & $0.1388$
    & $0.0872$ & $0.1706$ & $0.2458$ & $0.2599$ \\ 
    $Q_1(\pi/2)$\cite{pryadko-sengupta-2008} &${\pi/2}$ & $0$ & $-0.2906$ & $0$ & $-0.0023$ &
    $0.2021$ & $0.0040$ & $-0.0508$ \\ 
    $Q_2(\pi/2)$\cite{pryadko-sengupta-2008} &${\pi/2}$&  $0$ & $-0.0109$ & $0$ & $0.0035$ &
    $0.1617$ & $0.0730$ & $0.00077$ \\ 
    $W_{11}(\pi/2)$ &${\pi/2}$ &  $0$ & $0$ & $0.0106$ &
    $-0.0022$ & $0.1787$ & $0.0114$ & $0.0193$ \\ 
    $W_{12}(\pi/2)$ &${\pi/2}$ & $0$ & $0$ & $-0.0057$ &
    $-0.0020$ & $0.1756$ & $0.0482$ & $0.0103$ \\ 
    $W_{21}(\pi/2)$ &${\pi/2}$ &  $0$ & $0$ & $0$ & $-0.0059$ &
    $0.1796$ & $0.0301$ & $0.0190$ \\ 
    $W_{22}(\pi/2)$ &${\pi/2}$ & $0$ & $0$ & $0$ & $-0.0021$ &
    $0.1771$ & $0.0324$ & $0.0129$ \\ 
    PKRU$_1(\pi/2)$\cite{Pasini-2009} & $\pi/2$ & $0$ & $-0.2529$ & $-0.0260$ &
    $-0.0242$ & $0.1992$ & $0.0541$ & $-0.0422$ \\
    PKRU$_2(\pi/2)$\cite{Pasini-2009} & $\pi/2$ & $0$ & $0.2711$ & $0$ & $0.0127$ &
    $0$ & $0.0483$ & $0.00028$ 
  \end{tabular}
  \caption{Expansion coefficients for different pulse shapes.  See
    text for the definitions.}
  \label{tab:pulses}
\end{table*}

The first-order average evolution operator, Eq.~(\ref{eq:G1}), contains the
double integral of the commutator of the rotating-frame evolution operator
$\hat \Gamma(t_i)$ taken at different moments $t_i$, $i=1,2$.  There are only
a few combinations of trigonometric functions of the symmetrized
angle~(\ref{eq:symmetrized-angle}) and the time-independent terms that result
in non-zero contributions to $\overline\G^{(1)}$ for a single symmetric pulse
with the duration $\tau_p$:
\begin{eqnarray}
  \label{eq:alpha}
  \alpha&\equiv& \langle\sin(\varphi-\varphi')\rangle,\\
  \label{eq:zeta}
  \zeta&\equiv& \left\langle \left({t\over \tau_p}-{1\over 2}\right)\sin
    \varphi\right\rangle,\\
  \alpha_2&\equiv& \langle\sin(2\varphi-2\varphi')\rangle,\\
  \label{eq:zeta2}
  \zeta_2&\equiv& \left\langle \left({t\over \tau_p}-{1\over 2}\right)\sin
    2\varphi\right\rangle,\\
  \label{eq:mu}
  \mu&\equiv&\langle\sin(2\varphi-\varphi')\rangle,
\end{eqnarray}
where $\varphi\equiv \varphi(t)$, $\varphi'\equiv \varphi(t')$, and
the two-parameter averages are defined as
\begin{equation}
  \label{eq:two-param-average}
  \langle f(t,t')\rangle\equiv {1\over \tau_p^2}\int_0^{\tau_p} dt\int_0^t
  dt'\, f(t,t').
\end{equation}
The parameters $\alpha$ and $\zeta$ completely characterize the Hermitian
evolution during the pulse; they were introduced and discussed in detail in
Refs.~\onlinecite{pryadko-quiroz-2007,pryadko-sengupta-2008}.

\subsubsection{Decoupling sequences of finite-length  $\pi$-pulses}

Given the computed terms of the average evolution operator for a
single finite-duration pulse along the $x$ axis, the corresponding
expressions for a pulse along an arbitrary direction can be found by
an orthogonal transformation, using the appropriately transformed
vector ${\bf B}$ and decoherence matrix $\hat\gamma$.  We checked that
in simple cases, $\hat {\bf n}=\pm {\bf e}_x$ and $\hat {\bf
  n}=\pm{\bf e}_y$, the results coinside identically with those obtained
directly as outlined in Sec.~\ref{sec:single-pulse}.  

With these expressions, the evolution matrix $\hat Q$ for a sequence
of pulses can be computed as a product of the evolution operators for
individual pulses,
$$\hat Q_i=\hat
Q_{0,i}
\Bigl[\openone-\tau_p\overline\G^{(0)}_i-\tau_p\overline\G^{(1)}_i
  +{\tau_p^2\over2} \overline\G^{(0)}_i\cdot
  \overline\G^{(0)}_i+\mathcal{O}(\tau_p^3)\Bigr].  
$$
For a sequence of pulses, the net average evolution operator $\overline
\Gamma_{\rm seq}$ can be obtained in terms of a logarithm of the obtained
series (multiplied from the left by the corresponding $\hat
Q_{0,\mathrm{seq}}$ when it is not equal to $\openone$).

For the cycle ${\bf 2s}=\X \X$ of the Carr-Purcell sequence, the
leading-order evolution operator is the same as for a single
$\pi_x$-pulse, Eq.~(\ref{eq:one-pi-pulse-evol0}), but with
$\upsilon\to0$.  The corresponding expression in the subleading order
is complicated and contains terms proportional to $\upsilon$,
$\upsilon^2$, $\upsilon\upsilon_2$,  
$\alpha$, $\alpha_2$, $\zeta_2$; all of these terms
disappear in the limit of hard pulses.  An antisymmetric version of
the same sequence, ${\bf 2a}=\bar \X \X$, results in the leading-order
evolution operator of the form~(\ref{eq:one-pi-pulse-evol0}),
including the terms with $\upsilon$, but the subleading order
disappears, $\overline\G^{(1)}({\bf 2a})=0$.  Finally, for the
phase-ramped four-pulse sequence ${\bf 4a}=\bar \X\bar \X\X\X$, the
leading-order 
effective decoherence operator 
is the same as for the Carr-Purcell cycle, 
$$\overline\G^{(0)}({\bf
  4a})=\overline\G^{(0)}({\bf 2s})=\overline\G^{(0)}(\X)\Bigr|_{\upsilon\to0}, 
$$
while the subleading order vanishes, $\overline\G^{(1)}({\bf 4a})=0$. 

For the two-dimensional decoupling sequence ${\bf 4p}=\X\Y\bar\X\Y$, the
leading-order effective decoherence operator reads 
\begin{widetext}
  \begin{equation}
    \overline\G^{(0)}({\bf 4p})=\left(
      \begin{array}{ccc}
        \gamma_{yy}+\gamma_{zz} {3-\upsilon_2\over4}+\gamma_{xx}{
          1+\upsilon_2\over4} & - {\upsilon\over2}( B_x+  \gamma_{yz})
          &  {\upsilon\over2}( \gamma_{xy}-      B_z) \\
          {\upsilon\over2}( B_x- \gamma_{yz}) &  \gamma_{xx}+\gamma_{zz}
          {3-\upsilon_2\over4}+\gamma_{yy} {1+\upsilon_2\over4} & 0 \\
          { \upsilon\over2} (B_z+ \gamma_{xy}) & 0 & (\gamma_{xx}+\gamma_{yy})
          {3-\upsilon_2\over4}+ \gamma_{zz} 
          {1+\upsilon_2\over2}
        \end{array}
      \right)\label{eq:leading-order-4p}
    \end{equation}
\end{widetext}
Note that the corresponding diagonal terms are determined by the
diagonal terms of the original matrix $\hat\gamma$ and the parameter
$\upsilon_2$ [see Eq.~(\ref{eq:avs-1st})], while off-diagonal terms
are proportional to the effective width of the pulse $\upsilon$.  For
Hamiltonian evolution, $\hat\gamma=0$, one could achieve
$\overline\Gamma^{(0)}(\mathbf{4p})=0$ by using self-refocusing pulses
with $\upsilon=0$.  The next order term is complicated, but it is
eliminated for the antisymmetric sequence, ${\bf
  8a}=\X\Y\bar\X\Y\bar\Y\X\bar\Y\bar\X$.  Specifically, for this
sequence we obtain the same leading-order expression
$\overline\G^{(0)}({\bf 8a})=\overline\G^{(0)}({\bf 4p})$
[Eq.~(\ref{eq:leading-order-4p})], and $\overline\G^{(1)}({\bf
  8a})=0$.  On the other hand, for the symmetric sequence ${\bf
  8s}=\X\Y\bar\X\Y\Y\bar\X\Y\X$, it is the leading-order off-diagonal
corrections proportional to $\upsilon$ that vanish,
$$\overline\G^{(0)}({\bf 8s})=\overline\G^{(0)}({\bf
  4p})\bigr|_{\upsilon\to0},
$$
while the subleading order contains terms proportional to $\upsilon$,
$\alpha$, $\alpha_2$, and $\zeta_2$.  All of the corrections can be eliminated
by a supercycle of phase ramping which leads to an antisymmetric 16-pulse
sequence {\bf 16a} constructed as the {\bf 8s} sequence followed by the same
sequence with inverted pulses.  Specifically,
\begin{eqnarray}
  \label{eq:16a-0th}
\overline\G^{(0)}({\bf 16a})&=&\overline\G^{(0)}({\bf
  8s})=\overline\G^{(0)}({\bf 8a})\Bigr|_{\upsilon\to0},\\
  \label{eq:16a-1st}
\overline\G^{(1)}({\bf 16a})&=&\overline\G^{(1)}({\bf
  8a})=0,
\end{eqnarray}
where $\overline\G^{(0)}({\bf 8a})=\overline\G^{(0)}({\bf 4p})$ is given by
Eq.~(\ref{eq:leading-order-4p}).   We should mention that for Hamiltonian
evolution, $\hat\gamma=0$, one could achieve $\overline\G^{(0)}({\bf
  8s})=\overline\G^{(1)}({\bf 8s})=0$ by using second-order pulses with
$\upsilon=\alpha=0$ \cite{pryadko-quiroz-2007,pryadko-sengupta-2008}.

Comparing these results with the analogous sequences of hard pulses,
we see that an equivalent cancellation with soft pulses requires 
careful pulse shaping or doubling the number of pulses in the
sequence.  On the other hand, while sequences of hard $\pi$ pulses do
not modify the diagonal part of the effective decoherence operator,
this is not so with shaped pulses where $\upsilon_2\neq0$.  This
corresponds to a {\em reduction\/} of the effective decoherence rate
$\overline T_1^{\,-1}$ due to redistribution between the channels.  For
example, with the NMR decoherence model, Eq.~(\ref{eq:Gamma-phase}),
the inverse decoherence time becomes
\begin{equation}
  \label{eq:decoherence-time-4p}
  \overline T_1^{\,-1}=2\gamma-(\gamma-\gamma_\phi){1+\upsilon_2\over2};
\end{equation}
for $\gamma>\gamma_\phi$ and limited control fields
($|\upsilon_2|<1$), this is smaller than the original
$T_1^{-1}=2\gamma$.  Note that the reduction is achieved solely by
redistribution of decoherence between the channels, that is, by a
corresponding increase in $\overline T_2^{\,-1}$.  For full
symmetrization [Eq.~(\ref{eq:Gamma-symmetric})] one needs
$\upsilon_2=1/3$.

\subsubsection{Sequences of finite-length $\pi/2$ pulses}

For the family of sequences {\bf 12}, {\bf 24}, {\bf 48} [see
  Eqs.~(\ref{eq:12}) -- (\ref{eq:48})] of $\pi/2$ pulses, the results
are similar to those with hard pulses.  Specifically, when used with
NMR decoherence model~(\ref{eq:Gamma-phase}), the sequence {\bf 12}
achieves leading-order coherence matrix symmetrization but not the
decoupling.  The symmetrized (and phase ramped) sequences {\bf 24} and
{\bf 48} achieve symmetrization and decoupling in the leading and
subleading orders, respectively.  More explicitly, for symmetric
  pulse shapes and the decoherence model~(\ref{eq:Gamma-phase}), 
\begin{equation}
  \label{eq:half-pi-results}
  \overline\G^{(0)}({\bf 24)}=\overline\G^{(0)}({\bf
  48)}={2\over3}(2\gamma+\gamma_\phi)\openone,\;\, \overline\G^{(1)}({\bf
  48)}=0.
\end{equation}
The non-zero matrix elements in $\overline\G^{(1)}({\bf 24)}$ contain terms
scaling as the products of $(\gamma-\gamma_\phi)$ and the external field
components $B_\mu$ in all combinations; all of the coefficients cannot be
eliminated merely by pulse shaping.  In the special case $B_\mu=0$, the
correction $\overline\G^{(1)}({\bf 24)}\propto (\gamma-\gamma_\phi)^2$ is
suppressed for pulse shapes with $\upsilon_2=\alpha_2=0$.

With generic decoherence matrix $\hat\gamma$,
$\overline\G^{(1)}(\mathbf{48})=0$ for $90$-degree pulses with
$\upsilon=\upsilon_2=0$.  In this case the full symmetrization of the
decoherence matrix can be achieved as long as $\gamma_{xx}=\gamma_{yy}$, which
can be always made to be the case by an appropriate choice of the basis in the
$x$--$y$ plane.

\subsection{Pulse shaping}
The obtained analytical results imply two possible applications for pulse
shaping.  \textbf{First}, a pulse can be shaped to have the coefficients
$\upsilon$, $\alpha$, $\alpha_2$, and $\zeta_2$ zero, so that the
second-order correction to effective decoherence operator is zero already for
the 8-pulse sequence \textbf{8s}.  The resulting effective decoherence
operator $\overline\G$ will be diagonal, with the matrix elements determined
by the diagonal terms of the original decoherence matrix
$\hat\gamma$ and the parameter $\upsilon_2$ [cf.\ the diagonal elements in
Eq.~(\ref{eq:leading-order-4p})].  We note that up to terms of second order in
pulse duration, one might as well use the longer sequence \textbf{16a} to
cancel the terms in $\overline\G$ proportional to coefficients $\upsilon$,
$\alpha$, $\alpha_2$, $\zeta_2$.

The value $\upsilon_2=-1$ corresponds to an ideal hard inversion
pulse, in which case the diagonal matrix elements of the matrix
$\overline\G$ are not modified.  On the other hand, for
$\upsilon_2=1/3$, assuming $\gamma_{xx}=\gamma_{yy}$ (which can always
made to be the case by a rotation in the $x$--$y$ plane), the
effective decoherence operator is fully symetrized [see
  Eq.~(\ref{eq:Gamma-symmetric})].  This is the \textbf{second}
potential application for pulse shaping.  Again, to second order, the
same symmetrization can be also achieved with the composite-pulse
sequence \textbf{48} using $\pi/2$-pulses with $\upsilon=\upsilon_2=0$
(generic decoherence matrix), or any pulses for the special case of
NMR decoherence model~(\ref{eq:Gamma-phase}).

\begin{figure}[htbp]
  \centering
  \epsfxsize=\columnwidth
  \epsfbox{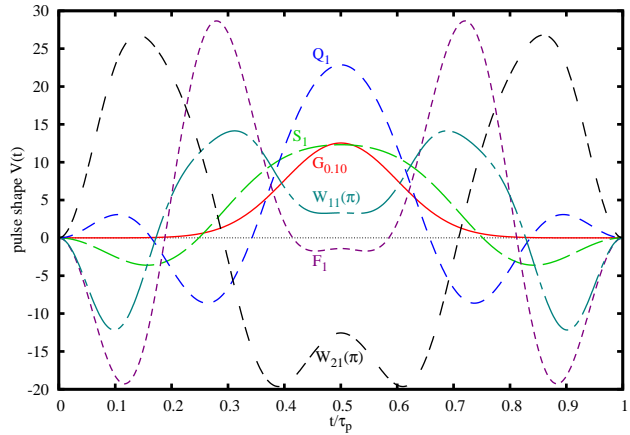}
  \caption{(Color online) Pulse shapes used in the simulations.  The
  Gaussian shape $G_{x}(y)=[(2\pi)^{1/2}x\tau_p]^{-1} 
  \exp(-0.5 \,y^2/x^2)$
  with $x=0.10$ and $y=t/\tau_p-0.5$ is $\sqrt2$ wider than that used in
  Ref.~\cite{pryadko-quiroz-2007}.  The shapes $S_1$, $Q_1$ were
  originally constructed in Ref.~\cite{sengupta-pryadko-ref-2005}.
  The shapes $F_1$, $W_{11}$, $W_{21}$ are constructed in this work,
  see Tab.~\ref{tab:coeff} for the coefficients.}
  \label{fig:pulses}
\end{figure}

We followed
Refs.~\onlinecite{sengupta-pryadko-ref-2005,pryadko-sengupta-2008} to
construct a number of new $\pi/2$ and inversion ($\phi_0=\pi$) pulse
shapes with $\upsilon=\upsilon_2=0$ [$W_{1s}(\phi_0)$],
$\upsilon=\upsilon_2=\alpha=0$ [$W_{2s}(\phi_0)$], and
$\upsilon=\upsilon_2=\alpha=\zeta_2=0$ [$W_{3s}(\phi_0)$], where the
parameter $s=1,2$ determines the number of derivatives that vanish
at the ends of the pulse interval (2 and 4, respectively).  The
parameters of the new and previously constructed pulses are listed in
Tab.~\ref{tab:pulses}, and their coefficients in Tab.~\ref{tab:coeff}.
The shapes of the pulses we used in the simulations are shown in
Fig.~\ref{fig:pulses}.

\begin{table*}[htbp]
  \centering
  \begin{tabular}[c]{c|c|c|c|c|c|c|c|c}
    pulse 
    & A0 & A1 & A2 & A3 & A4 & A5 & A6 & A7  \\ \hline
    $F_1$ & 0.5 & -1.419474 &  -2.048028 &
    1.549555 & 1.435813 & -0.017867 & \\ 
 $W_{11}(\pi)$ & 0.5&   -1.242022&   -1.009075 &   0.700828 &
    0.530624 & 1652161644 &  0.277982 &   0.241663 \\ 
 $W_{12}(\pi)$ &  0.5 &   -1.291342 &  -0.753726 &  1.499438 &
    0.364546 &  0.012680 & -0.069983 & -0.261614 \\
 $W_{21}(\pi)$ &  0.5&  3.056086&  -1.295369&  -1.689687&  -0.062202&  -0.366646&  -0.142183\\
 $W_{22}(\pi)$ &  0.5& 2.776007& -2.473314& -1.782314& 0.958211&
    -0.444991& 0.300165& 0.166236\\ 
 $W_{31}(\pi)$ &  0.5&  -1.110710&  -3.692547&  1.248118&  0.990698&  1.394824&  0.669618\\
 $W_{32}(\pi)$ &  0.5&  -1.686664&  -2.108402&  3.362253&  1.029286&  -0.260405&  -0.836068\\
    \hline 
 $W_{11}(\pi/2)$ & 0.25& 2.011311& 0.041292& 1.381531& 0.262448& 0.076040&\\
 $W_{12}(\pi/2)$ & 0.25& 2.023581& 0.920572& 1.341484& -0.113434& -0.144034& -0.231008\\
 $W_{21}(\pi/2)$ & 0.25& 2.018463& 0.588295& 1.393403& -0.206226& 0.095943& -0.1029524&\\
 $W_{22}(\pi/2)$ & 0.25& 2.018283& 0.608538& 1.386685& 0.088935&
    0.024615& -0.134584& -0.205904
  \end{tabular}
  \caption{Pulse coefficients for various shapes constructed for this
    work.  The control field during the pulse, $0<t<\tau_p$, is represented as 
  $V(t)=2\pi\sum_n A_n \cos (2\pi n t/\tau_p)$.}
  \label{tab:coeff}
\end{table*}

\section{Simulations}
\label{sec:simulations}

Our analytical analysis was only done to second order cumulant
expansion, and assuming time-independent fields $B_\mu$.  The results
are asymptotically true in the limit where the sequences are
sufficiently short, that is, $\tau B\ll1$, $\gamma\tau\ll1$, $\tau\ll
\tau_c$, where $\tau_c\sim \omega_c^{-1}$ is the correlation time of
the slow part of the bath.  To check for possible effect of ignored
terms, we performed numerical simulations of the controlled dynamics
of a qubit in the presence of time-dependent correlated classical
Gaussian fields $B_\mu\equiv B_\mu(t)$, as well as Markovian
decoherence described by the Lindblad equation~(\ref{eq:Lindblad}),
(\ref{eq:decoherence}).  We specifically used the NMR decoherence
model~(\ref{eq:Gamma-phase}) in the regime dominated by phase
fluctuations, $\gamma_\phi>0$, $\gamma=0$ [note that with a non-zero
  $\gamma$ the decoherence matrix can be decomposed as
  $\hat\gamma=\gamma\openone+(\gamma_\phi-\gamma)\diag(0,0,1)$; the
  part proportional to the identity matrix commutes with decoupling
  pulses and can be trivially eliminated.]  The fields $B_\mu(t)$ were
chosen as zero-average correlated Gaussian random functions with the
correlation function
\begin{equation}
  \langle B_\mu (t) B_\nu(t')\rangle=\delta_{\mu\nu}\,B_0^2 g(t-t'),\quad
  g(t)=e^{-t^2/2 \tau_c^2}, 
  \label{eq:B-corr}
\end{equation}
and the fixed value of the correlation time $\tau_c=8\tau_p$.  The
control fields $V_\mu(t)$, $\mu=x,y$, were generated according to the
chosen decoupling sequence and the pulse shape, with the $k$-th pulse
of the sequence fitting in the interval $(k-1)\tau_p \le t\le
k\tau_p$; the corresponding decoupling period $\tau=n\tau_p$ is given by
the number $n$ of pulses in the sequence.  For every sequence, pulse
shape, and the realization of the random fields $B_\mu(t)$, we solved
Eqs.~(\ref{eq:Revol}), (\ref{eq:Gamma}) with $\vec\mathcal{R}=0$ for three
sets of initial conditions with $R_\mu=1$, $\mu=x,y,z$; the solutions
correspond to the columns of the evolution matrix $\hat Q(t)$
[Eq.~(\ref{eq:uniform-sol})].  The  average decoupling
fidelity (\ref{eq:av-fid}) was then evaluated using the expression
\begin{equation}
  \label{eq:fidelity-Q}
   \langle
F(t_s)\rangle={1\over2}\left(1+{\tr\hat Q(t_s)\over3}\right).
\end{equation}
at the time moments $t_s$ commensurate with the sequence duration
$t_s=s\tau=s n\tau_p$.

\begin{figure}[htbp]
  \centering
  \epsfxsize=\columnwidth
  \epsfbox{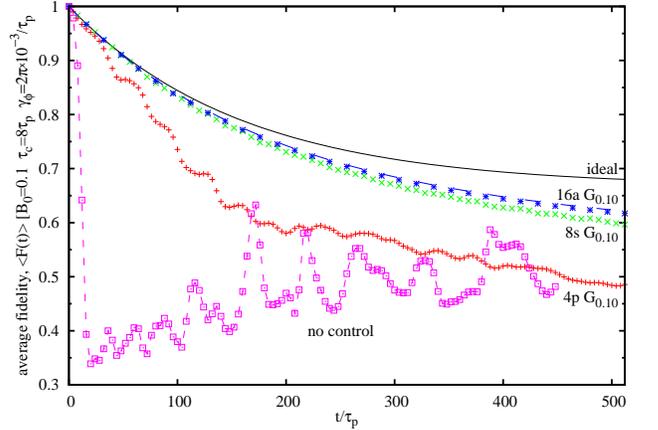}
  \caption{(Color online) Single-qubit average fidelity for a
    particular realization of time-dependent fields and Markovian
    dephasing with the rate $\gamma_\phi=2\pi\times 10^{-3}/\tau_p$.
    Open boxes connected with the dashed line marked ``no control''
    correspond to the absence of decoupling.  Other symbols are
    numerical data for the indicated decoupling sequences of Gaussian
    inversion pulses.  The black solid line marked ``ideal''
    corresponds to maximum achievable average fidelity.  Dashed line
    corresponds to Eq.~(\ref{eq:2nd-loss}) with $\upsilon_2=-0.7086$
    as appropriate for the pulse $G_{0.10}$ [see
      Tab.~\ref{tab:pulses}].}
  \label{fig:samp-G010}
\end{figure}

The calculation results for a particular sample of the fields
$B_\mu(t)$ and several sequences of inversion ($\phi_0=\pi$) Gaussian
pulses are shown in Fig.~\ref{fig:samp-G010}.  The line marked as
``ideal'' corresponds to the best achievable fidelity
\begin{equation}
  \langle F(t)\rangle_{\text{ideal}}={1\over3}(2+e^{-\gamma_\phi  t});
\label{eq:best-fidelity}
\end{equation}
with the chosen value $\gamma_\phi=2\pi\times 10^{-3}/\tau_p$ at the
end of the simulation we get $\langle F(512\tau_p)\rangle_{\rm ideal}\approx
0.680$.  The points marked ``no control'' corresponds to average
fidelity in the absence of decoupling pulses; the coherence is lost
after just a few pulse durations.  While decoupling is not
particularly efficient (and noise is evident) for the 4-pulse sequence
(set of points marked \textbf{4p} in Fig.~\ref{fig:samp-G010}), the
fidelity is improved and noise markedly reduced for the sequence
\textbf{8s}, and even more so for the sequence \textbf{16a}.  In fact,
the points for sequence \textbf{16a} are very close to the dashed line
which represents the leading-order contribution at $B=0$, see
Eq.~(\ref{eq:2nd-loss}) below.

\begin{figure}[htbp]
  \centering
  \epsfxsize=\columnwidth
  \epsfbox{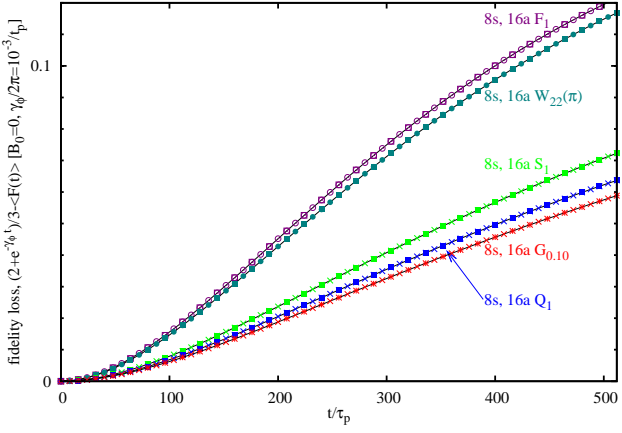}
  \caption{(Color online) Single-qubit average fidelity loss due to
    redistribution of the decoherence rates over directions in the
    absence of low-frequency fields.  Symbols are numerical data for
    pulses and sequences as indicated, and lines are computed
    analytically using Eq.~(\ref{eq:2nd-loss}) with corresponding
    $v_2$ from Tab.~\ref{tab:pulses}.  This contribution to infidelity
    would be zero for hard ($\delta$-function) pulses with
    $\upsilon_2=-1$.}
  \label{fig:2nd-loss}
\end{figure}

In the presence of the Markovian decoherence, we can view the average
infidelity as composed of three terms.  {\bf First} is the inavoidable
infidelity due to the Markovian decoherence alone, in the absence of
any control fields or low-frequency noise.  For the case of pure
dephasing, it is given by $1-\langle F(t)\rangle_{\rm ideal}$, see
Eq.~(\ref{eq:best-fidelity}).  For the value of $\gamma_\phi$ used in
our simulation, $\langle F(t)\rangle_{\rm ideal}$ is shown in
Fig.~\ref{fig:samp-G010} with the solid line.  {\bf Second} is the
fidelity loss due to the redistribution of the original decoherence
rate(s) over the channels.  In our simulations, for the sequences {\bf
  4p}, {\bf 8s}, {\bf 16a} of $\pi$-pulses other than $G_{0.10}$, this
part of the average infidelity is to a very good accuracy determined
by the coefficient $\upsilon_2$ of the corresponding pulses,
\begin{eqnarray}
  \label{eq:2nd-loss}
  \langle F(t)\rangle_0&=&{1\over6}\left(3+e^{-\gamma_{1\,\rm eff}t}+
    2e^{-\gamma_{2\,\rm eff} t}\right),\\
   \gamma_{1\,\rm eff}&=& \gamma_\phi{1+\upsilon_2\over2},\;
  \gamma_{2\,\rm eff}=\gamma_\phi{3-\upsilon_2\over4},
\end{eqnarray}
see Tab.~\ref{tab:pulses} and Fig.~\ref{fig:2nd-loss}.  

Finally, the {\bf third} part of the average infidelity is entirely
due to the presence of the low-frequency random fields; it is defined
as the difference between the average fidelity of the controlled
system in the absence of low-frequency fields, and that in their
presence, $\Delta F(t)\equiv \langle F(t)\rangle_{0}-\langle
F(t)\rangle$.  It is this quantity that directly characterizes the
ineffectiveness of the decoupling against the low-frequency fields.

\begin{figure}[htbp]
  \centering \epsfxsize=\columnwidth 
  \epsfbox{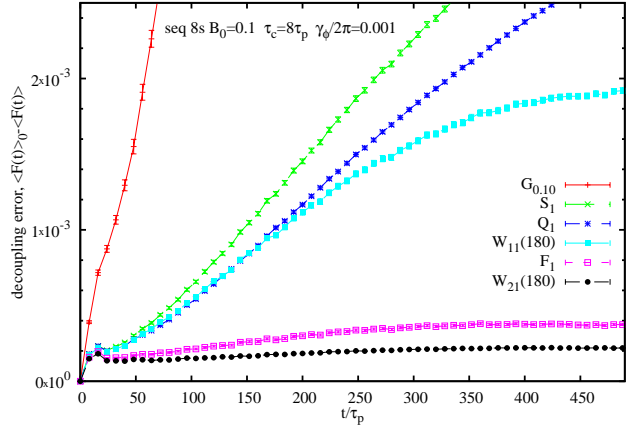}
  \caption{(Color online) Single-qubit decoupling error  for the
  8-pulse sequence 
    {\bf 8s} with several pulse shapes, as indicated.  The decoupling
    error is small for the second-order pulse $W_{21}(\pi)$ and,
    somewhat surprisingly, for the symmetrizing shape $F_1$.  With
    $\gamma_\phi=0$, the infidelity for this pulse shape with sequence
    {\bf 8s} grows similarly to that of first-order pulses $S_1$ and
    $W_{11}$ [Fig.~\protect\ref{fig:df-8s-nog}].}
  \label{fig:df-8s}
\end{figure}

\begin{figure}[htbp]
  \centering
  \epsfxsize=\columnwidth
  \epsfbox{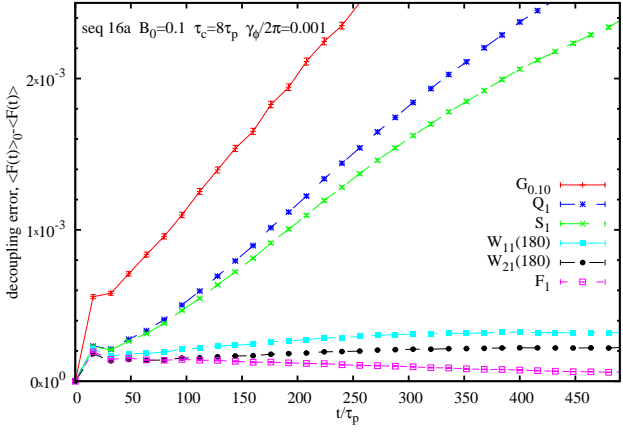}
  \caption{(Color online) As in Fig.~\ref{fig:df-8s} but for 16-pulse
    sequence {\bf 16a}.  For this sequence, and with time-independent
    fields $B_\mu$, there would be no contribution from the first two
    orders of the effective decoherence operator
    [Eqs.~(\protect\ref{eq:16a-0th}), (\protect\ref{eq:16a-1st})]; the
    decoupling error is determined by the non-adiabaticity and by the
    terms of higher order.  Generic pulse shapes $G_{0.10}$, $S_1$,
    and $Q_1$ show the largest decoupling error, while it is the
    smallest for the symmetrizing shape $F_1$ (technically, zeroth
    order, $\upsilon\neq0$; $\upsilon_2\approx 1/3$).  For shapes
    $F_1$, $W_{11}$, and $W_{21}$, the decoupling error here is
    actually smaller than the infidelity in
    Fig.~\protect\ref{fig:df-16a-nog}.  We believe this is an artefact
    of the definition of the decoupling error related to the full
    average fidelity approaching saturation at 1/2
    [Fig.~\ref{fig:samp-G010}].}
  \label{fig:df-16a}
\end{figure}

\begin{figure}[htbp]
  \centering
  \epsfxsize=\columnwidth
  \epsfbox{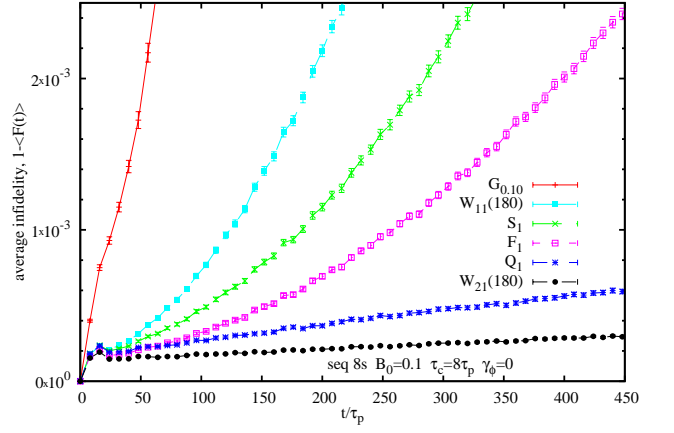}
  \caption{(Color online) As in Fig.~\ref{fig:df-8s} but in the
    absence of Markovian dephasing, $\gamma_\phi=0$.  Infidelity
    curves with 2nd-order pulses $Q_1$ and $W_{21}$ have a finite
    slope, unlike the plots in
    Ref.~\protect\onlinecite{pryadko-sengupta-kinetics-2006} where a
    smaller r.m.s.\ value $B_0$ of the slow fields was used.  We
    checked that the slopes for the 2nd-order pulse shapes are
    greately reduced if $B_0$ is reduced by a factor of two (not
    shown).}
  \label{fig:df-8s-nog}
\end{figure}

\begin{figure}[htbp]
  \centering
  \epsfxsize=\columnwidth
  \epsfbox{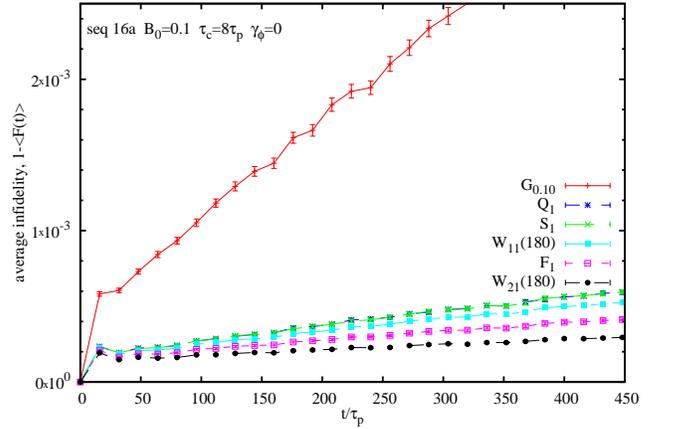}
  \caption{(Color online) As in Fig.~\ref{fig:df-8s-nog} but for the
    sequence {\bf 16a}.  The slopes for all the pulse shapes excluding
    $G_{0.10}$ are greately reduced if $B_0$ is reduced by a factor of
    two (not shown).  For the Gaussian pulses $G_{0.10}$, the
    second-order cancellation is achieved only after the complete
    sequence with the duration $\tau=16\tau_p$; the slope reduction
    happens for slower fields $B_\mu(t)$ with $\tau_c=16\tau_p$ (not
    shown).}
  \label{fig:df-16a-nog}
\end{figure}

The corresponding plots are shown in Figs.~\ref{fig:df-8s} (sequence
{\bf 8s}) and \ref{fig:df-16a} (sequence {\bf 16a}), as well as in
Figs.~\ref{fig:df-8s-nog}, \ref{fig:df-16a-nog} with $\gamma_\phi=0$
for comparison purposes.  The general trend is consistent with the
expectations based on analytical expansion.  For example, with the
Gaussian pulses we get the largest error, with or without the
Markovian dephasing; the decoupling error is smaller with the {\bf
  16a} sequence [as compared with the {\bf 8s}] where the corrections
including the subleading order are suppressed.  Similarly, specially
designed pulse shapes with a larger number of suppressed parameters
lead to improved performance in the presence of Markovian dephasing:
compare, e.g., the curves for pulses $S_1$, $W_{11}(\pi)$, and
$W_{21}(\pi)$.

One marked exception is the pulse shape $F_1$, for which
$\upsilon_2\approx+1/3$ while the other coefficients have magnitude
comparable to those for, e.g., the shape $S_1$ (apart from the
coefficient $\upsilon$ which is zero for the shape $S_1$).  While in
the absence of Markovian dephasing, the decoupling error for the pulse
$F_1$ is qualitatively similar to that for the 1st-order shapes
[Figs.~\ref{fig:df-8s-nog}, \ref{fig:df-16a-nog}], with
$\gamma_\phi\neq0$ the decoupling error is actually closer to that for
the {\em second\/}-order pulses with the sequence {\bf 8s}
[Fig.~\ref{fig:df-8s}] and is the smallest of the shapes we considered
for the sequence {\bf 16a} [Fig.~\ref{fig:df-16a}].  We believe that
this, at least in part, is associated with the symmetrization of the
effective decoherence tensor achieved in the presence of the pulse
$F_1$, which renders the resulting Markovian decoherence (effectively,
the amplitude damping) independent from the decoupling, and thus
reduces the contribution from high orders not included in our
analytical calculations.


In comparison with the sequences of $\pi$-pulses, the basic unit in
the sequences of $\pi/2$ pulses, {\bf 12p}, {\bf 24s}, and {\bf 48s},
is three times longer.  As a result, weaker random fields $B_\mu(t)$
with longer correlation times are required to achieve comparable
decoupling accuracy.  In addition, with longer coherence time, longer
evolution times are required to separate the effects of the transients
near the beginning of the simulation interval.  By these reasons, we
do not discuss the performance of these sequences in detail.  For the
parameters as in Figs.~\ref{fig:df-8s}, \ref{fig:df-16a}, the
decoupling error with the sequences {\bf 24s} and {\bf 48s} saturates
at around twenty times that for the sequences {\bf 8s}, {\bf 16a},
while with $\tau_c=32\tau_p$ and the same $B_0=0.1/\tau_p$, the
decoupling error is comparable to that in Figs.~\ref{fig:df-8s},
\ref{fig:df-16a}.  Note that these decoupling errors are still small
compared to the fidelity loss due to redistribution of the decoherence
rates over directions, which for these sequences is identical to that
of sequences {\bf 8s}, {\bf 16a} with the symmetrizing pulse $F_1$
[see Fig.~\ref{fig:2nd-loss}].

\section{Conclusions}

We considered the dynamical decoupling in a simple decoherence model
simulating the presence of both low- and high-frequency environment
modes [Fig.~\ref{fig:spectr}], having in mind applications of DD in
combination with QECCs.  We modeled the effect of low-frequency
degrees of freedom in terms of classical correlated noise, and that of
the fast degrees of freedom with the help of the Markovian master
equation in the Lindblad form, see Eqs.~(\ref{eq:Lindblad}),
(\ref{eq:dissipator}).  The combined effect is the non-Hermitian
evolution (\ref{eq:Revol}) of the Bloch vector with the instantaneous
decoherence operator~(\ref{eq:Gamma}) [see Eq.~(\ref{eq:Gamma-phase})
  for the special case of NMR decoherence model].  In the presence of
dynamical decoupling with the sequence period $\tau$, the values of
the average qubit fidelity at the commensurate time moments
$\tau_s=s\tau$ are determined by the effective decoherence matrix
$\overline\Gamma$ [Eq.~(\ref{eq:Sformal})], an analogue of the average
Hamiltonian.  The trace of the effective decoherence matrix is
conserved, see Eq.~(\ref{eq:trace}); in agreement with general
expectations, this implies that dynamical decoupling can only decrease
the fidelity in the presence of Markovian degrees of freedom alone.

This is always the case for DD with soft pulses, which necessarily
leads to redistribution of the decoherence rates over the directions,
leading to some fidelity reduction even in the absence of the slow
degrees of freedom.  For evolution time small compared to the
decoherence time $\sim\gamma^{-1}$ (as required for efficient error
correction), this reduction is a relatively small effect.  On the flip
side, this redistribution causes symmetrization of decoherence
operator which reduces the effect of the non-Hamiltonian dynamics
associated with the fast degrees of freedom on the decoupling
accuracy.  For example, if the
symmetrization~(\ref{eq:Gamma-symmetric}) is achieved at the end of a
basic decoupling cycle, decoherence is expected to have no effect on
additional cancellations achieved in a supercycle obtained by, e.g.,
phase-ramping the basic cycle.

To analyze relative importance of these effects, we considered several
decoupling sequences of both hard and soft (generic 1st- and 2nd-order
and specially shaped) pulses which lead to various degrees of
symmetrization of the decoherence operator.  For such sequences, in
the static limit where the slow degrees of freedom become
time-independent, we constructed analytically the first two terms of
the (cumulant) expansion of the effective decoherence operator
$\overline\Gamma$ in powers of the sequence period $\tau$.  With the
help of pulse shaping and/or phase ramping, we could ensure that the
leading-order average decoherence operator $\overline\Gamma^{(0)}$ is
diagonal and independent of the components of the slow field, while
the subleading contribution disappears identically,
$\overline\Gamma^{(1)}=0$.  Results of
Refs.~\onlinecite{sengupta-pryadko-ref-2005,pryadko-sengupta-kinetics-2006}
suggest that such sequences might also be effective in the presence of
sufficiently weak slowly-varying fields.

Numerical simulations generally confirmed these expectations.  In the
studied parameter range, the best decoupling accuracy is achieved with
the sixteen-pulse sequence {\bf 16a}.  With time-independent external
fields the sequence provides complete decoupling of the slow fields in
the first two orders of the cumulant expansion independent of the
shape of symmetric inversion pulses, see Eqs.~(\ref{eq:16a-0th}),
(\ref{eq:16a-1st}).  The effects of higher order terms and
non-adiabaticity of the fields $B_\mu(t)$ are captured in the
simulations, see Sec.\ \ref{sec:simulations}.  The best performance is
achieved with the ``symmetrizing'' pulse shape $F_1$ constructed with
the only requirement that the diagonal part of the leading-order
effective decoherence matrix~(\ref{eq:leading-order-4p}) for the
4-pulse sequence {\bf 4p} is symmetric, i.e., $\upsilon_2\approx
+1/3$, see Sec.~\ref{sec:single-pulse} and Tab.~\ref{tab:pulses}.  In
fact, in our simulations the part of the coherence loss associated
with the slow fields is smaller than that in the absence of the
high-frequency degrees of freedom, cf.~Figs.~\ref{fig:df-8s},
\ref{fig:df-16a} and \ref{fig:df-8s-nog}, \ref{fig:df-16a-nog}.
Overall, the suppression of the decoherence due to slow fields
associated with the dynamical decoupling increases the coherence time
by orders of magnitude.

We thus demonstrated in principle the effectiveness of DD in
suppressing the effect of low-frequency degrees of freedom in the
presence of high-frequency modes which cannot be eliminated by
dynamical decoupling.  This opens up possible applications in combined
coherence protection techniques concatenating dynamical decoupling
with QECC at higher levels.  We postpone the corresponding discussion
to a futher publication\cite{Li-Lidar-Pryadko-unpublished}.

\section*{Acknowledgments}
We are grateful to Sasha Korotkov, Daniel Lidar, Len Mueller, and
G\"otz Uhrig for useful discussions.  This research was supported in
part by the NSF grant No.\ 0622242.

\end{document}